\documentclass[12pt,a4]{article}
\usepackage{graphicx}

\begin{document}
\begin{flushright} 
CU-TP/01-05
\end{flushright}
\vspace*{1cm}
\begin{center}
{\LARGE Mixing among light scalar mesons and \\
$L=1\ q\bar{q}$ scalar mesons}\\
\vspace{0.5cm}       
{\large T. Teshima\footnote{E-mail: teshima@isc.chubu.ac.jp}, I. Kitamura 
and N. Morisita\\
{\it Department of Applied Physics,  Chubu University }\\
{\it Kasugai 487-8501, Japan}}
\end{center}
\vspace{0.5cm}
\begin{abstract}
Following the re-establishment of the $\sigma(600)$ and the $\kappa(900)$, the 
light scalar mesons $a_0(980)$ and $f_0(980)$ together with the $\sigma(600)$ 
and the $\kappa(900)$ are considered as the chiral scalar partner of 
pseudoscalar nonet in $SU(3)$ chiral symmetry, and the high mass scalar mesons 
$a_0(1450)$, $K^*_0(1430)$, $f_0(1370)$ and $f_0(1710)$ turned out to be 
considered as the $L=1\ q\bar{q}$ scalar mesons. We assume that the high mass 
of the $L=1\ q\bar{q}$ scalar mesons is caused by the mixing with the light 
scalar mesons. For the structure of the light scalar mesons, we adopted the 
$qq\bar{q}\bar{q}$ model in order to explain the "scalar meson puzzle". The 
inter-mixing between the light scalar nonet and the high mass $L=1\ q\bar{q}$ 
nonet and the intra-mixing among each nonet are analyzed by including the 
glueball into the high mass scalar nonet.
\end{abstract}
\vspace{0.5cm}
\hspace*{1cm}PACS number(s): {11.30.Rd, 12.39.Mk, 14.40.-n}
\newpage
\setlength{\baselineskip}{0.33in}
\section{INTRODUCTION}
The $\sigma$ meson was introduced theoretically in connection with the linear 
sigma model (L$\sigma$M)\cite{SIGMAMODEL} and listed in the Particle Data 
Group (PDG) until 1972. This meson disappeared in PDG until 1996, because 
this resonance was not narrow and the broad resonance could not easily be 
distinguished from a background. 
In recent analyses \cite{ISHIDA} of $\pi\pi$ and $\pi K$ scattering phase 
shift, the scalar mesons, $\sigma(600)$ with $I=0$ and $\kappa(900)$ 
with $I=1/2$, have been observed. These analyses use the interfering 
Breit-Wigner amplitude and introduce a negative background phase of a hard 
core type.
\par
In connection with the re-establishment of these particles, they and already 
established iso-singlet $f_0(980)$ and iso-triplet $a_0(980)$ lower than 1 
GeV turned out to be considered as a chiral partner ($\sigma$-nonet) of the 
pseudoscalar ($\pi$-nonet) in $SU(3)$ chiral symmetry \cite{MISHIDA}. Many 
authors analyzed the $f_0(980)$ and $a_0(980)$ mesons using the $K\bar{K}$ 
molecule model \cite{ISGUR} or  $qq\bar{q}\bar{q}$ model \cite{JAFFE} rather 
than the above L$\sigma$M in order to explain the so-called "scalar meson 
puzzle" why the $f_0(980)$ degenerate to the $a_0(980)$ has so large 
$K\bar{K}$ decay width compared to the $\pi\pi$ decay width. Anyway, the light 
scalar mesons with masses lower than 1 GeV are considered as not the 
conventional $L=1\ q\bar{q}$ scalar nonet by many authors. The high mass scalar mesons $a_0(1450)$, $K_0^*(1430)$, $f_0(1370)$ and $f_0(1710)$ turn out to be 
classified to the conventional $L=1\ q\bar{q}$ $SU(3)$ nonet.
\par
However, here we encounter a puzzle: why $L=1\ q\bar{q}$ scalar mesons 
considered above have so high masses compared with other $L=1\ q\bar{q}$ 
$1^{++}$ and $2^{++}$ mesons. If $L\cdot S$ interaction is assumed, there must 
be satisfied a relation for masses of them: $m^2(2^{++})-m^2(1^{++})=
2(m^2(1^{++})-m^2(0^{++}))$. But, experimentally, this relation is not 
satisfied at all: $m^2(1^{++})$ is less than $m^2(0^{++})$ as $m_{a_2}=
1318{\rm MeV}$, $m_{a_1}=1230{\rm MeV}$, $m_{a_0}=1474{\rm MeV}$, and 
$m_{K_2^*}=1429{\rm MeV}$, $m_{K_1}=1339{\rm MeV}$, $m_{K_0^*}=
1412{\rm MeV}$, where we used the average value of $K_1(1^{-+})$ and 
$K_1(1^{++})$ for the $m_{K_1}$. In order to solve this puzzle, we assume 
an inter-mixing between the light scalar nonet and the $L=1\ q\bar{q}$ nonet
\cite{BLACK}. Because the difference between the mass of $m_{a_0}=
1474{\rm MeV}$ and the mass $1236{\rm MeV}$ predicted from the relation 
$m^2(2^{++})-m^2(1^{++})=2(m^2(1^{++})-m^2(0^{++}))$ is very large, the 
inter-mixing between the light scalar meson and the $L=1\ q\bar{q}$ meson is 
considered to be very strong. The strength of the mixing depends on the 
structure of scalar mesons. As seen later, although we assume the structure 
of the light scalar mesons to be $qq\bar{q}\bar{q}$, the mixing strength 
can be very large. 
\par
There are another important mixing between two isoscalar mesons in both light 
scalar nonet and $L=1\ q\bar{q}$ nonet. We call the mixing "intra-mixing". 
Only the intra-mixing is considered in many analyses of masses and decays of 
the light scalar nonet and the $L=1\ q\bar{q}$ nonet without considerations 
of the inter-mixing. However, as seen later, the strength of the inter-mixing 
is considered to be rather large compared with the strength of the 
intra-mixing. Furthermore, the existence of the scalar glueball is suggested 
in QCD \cite{GLUEBALL1} and the $f_0(1500)$ is considered to be the most 
probable candidate for scalar glueball \cite{GLUEBALL2}. Then the mixing 
between the $I=0$ $L=1\ q\bar{q}$ meson and the scalar glueball has to be 
considered, because the mixing between the $L=1\ q\bar{q}$ meson and the 
scalar glueball is considered to be rather large compared with the mixing 
among the $I=0$ mesons. Thus we will study the overall mixing; inter-mixing, 
intra-mixing and glueball mixing, in order to analyze the masses and decays 
of the light scalar mesons and $L=1\ q\bar{q}$ mesons. 
\par
In section 2, we will discuss about the fact that the strength of the 
inter-mixing is very large. In section 3, we discuss the structure of the 
light scalar mesons. In section 4, we will analyze the overall mixing; 
inter-mixing, intra-mixing and glueball mixing.  
\section{Strength of inter-mixing}
In this section, we estimate the strength of the inter-mixing between the 
light scalar mesons and the conventional $L=1\ q\bar{q}$ scalar mesons. For 
the estimation of the inter-mixing, it is necessary to know the masses before 
mixing of the scalar mesons. The mass values of the $2^{++}$ and $1^{++}$ 
mesons cited in Particl Data Group \cite{PARTICLE} are as follows; 
\begin{equation}
\left\{\begin{array}{l}
m_{a_2(1320)}=1318
{\rm MeV},\ \  m_{K_2^*(1430)}=1429{\rm MeV},\\  
\hspace{2cm}m_{f_2(1270)}=1275{\rm MeV},\ \  m_{f'_2(1525)}=1525{\rm MeV}, \\
m_{a_1(1260)}=1230{\rm MeV},\ \  m_{K_1(1270/1400)}=1339{\rm MeV},\\  
\hspace{2cm}m_{f_1(1285)}=1282{\rm MeV},\ \   m_{f_1(1420)}=1426{\rm MeV}, 
\end{array}\right.
\end{equation} 
where we averaged the $m_{K_1(1270)}=1273{\rm MeV}$ and $m_{K_1(1400)}=
1402{\rm MeV}$ in quadratic mass because they are considered to be mixed up. 
We plot the $L=1\ q{\bar q}$ $2^{++}$, $1^{++}$ and $0^{++}$ mass spectra in 
Fig.~1. Because the mass differences between $m_{a_2(1320)}$, $m_{f_2(1270)}$ 
and beween $m_{a_1(1260)}$, $m_{f_1(1285)}$ are very small, we consider the 
ideal mixing limit, neglect these mass differences and average these masses as
\begin{equation}
\left\{\begin{array}{l}
m_{a_2(1320)}=m_{f_2(1270)}=1297{\rm MeV},\ m_{K_2^*(1430)}=
1429{\rm MeV},\ m_{f'_2(1525)}=1525{\rm MeV}, \\
\\
m_{a_1(1260)}= m_{f_1(1285)}=1256{\rm MeV},\ m_{K_1}=1339{\rm MeV},\ 
m_{f_1(1420)}=1426{\rm MeV}. 
\end{array}\right.
\end{equation} 
If we assume the $L\cdot S$ force for $L=1\ q\bar{q}$ bound states, 
then the following well-known mass relation is obtained,  
\begin{eqnarray}
&&m^2(2^{++})-m^2(1^{++})=2(m^2(1^{++})-m^2(0^{++})).
\end{eqnarray}
From this relation, the masses of $L=1\ q\bar{q}$ $0^{++}$ mesons before 
inter-mixing denoted by 
$\overline{a_0(1450)}$, $\overline{f_0(1370)}$, $\overline{K^*_0(1430)}$ 
and $\overline{f_0(1710)}$ are 
estimated as follows;
\begin{equation}
\begin{array}{l}
{m}_{\overline{a_0(1450)}}={m}_{\overline{f_0(1370)}}=1236{\rm MeV},\\  
m_{\overline{K^*_0(1430)}}=1307{\rm MeV},\ \ \ m_{\overline{f_0(1710)}}
=1374{\rm MeV}
\end{array} 
\end{equation}
where the ideal (intra-)mixing mass relation $m_{\overline{f_0(1710)}}^2=
2m_{\overline{K^*_0(1430)}}^2-m_{\overline{a_0(1450)}}^2$ is used.  
\par
Because the mass ${m}_{\overline{a_0(1450)}}$ for $I=1$ $a_0(1450)$ before 
inter-mixing is 1236MeV and the mass for light scalar 
$I=1$ meson $a_0(980)$ is 985MeV, the ${m}_{\overline{a_0(980)}}$ for 
$a_0(980)$ before inter-mixing is estimated as 1271MeV using the mass 
relation for 2-body mixing ${m}^2_{a_0(1450)}-{m}^2_{\overline{a_0(1450)}}=
{m}^2_{\overline{a_0(980)}}-{m}^2_{a_0(980)}$. Similarly, the mass 
${m}_{\overline{\kappa(900)}}$ for $I=1/2$ $\kappa(900)$ before inter-mixing 
to be 1047MeV is obtained from the masses $m_{\kappa(900)}=900{\rm MeV}$, 
$m_{K^*_0(1430)}=1412{\rm MeV}$ and $m_{\overline{K^*_0(1430)}}=
1307{\rm MeV}$. The mass $m_{\overline{\sigma(600)}}$ for $\sigma(600)$ before 
inter-mixing is obtained as $760{\rm MeV}$ from the ideal (intra-)mixing 
mass relation $m_{\overline{\sigma(600)}}^2=
2m_{\overline{\kappa(900)}}^2-m_{\overline{a_0(980)}}^2$. 
\begin{equation}
\begin{array}{l}
{m}_{\overline{a_0(980)}}={m}_{\overline{f_0(980)}}=1271{\rm MeV},\\  
m_{\overline{\kappa_0(900)}}=1047{\rm MeV},\ \ \ m_{\overline{\sigma(600)}}
=760{\rm MeV}.
\end{array} 
\end{equation}
\par
We can easily estimate the strength of inter-mixing for the $I=1$ $a_0(1450)$ 
and $a_0(980)$ and $I=1/2$ $K_0^*(1430)$ and $\kappa(900)$ because these 
states have no effects from the intra-mixing. 
\begin{figure}
\begin{picture}(5,5)
\input{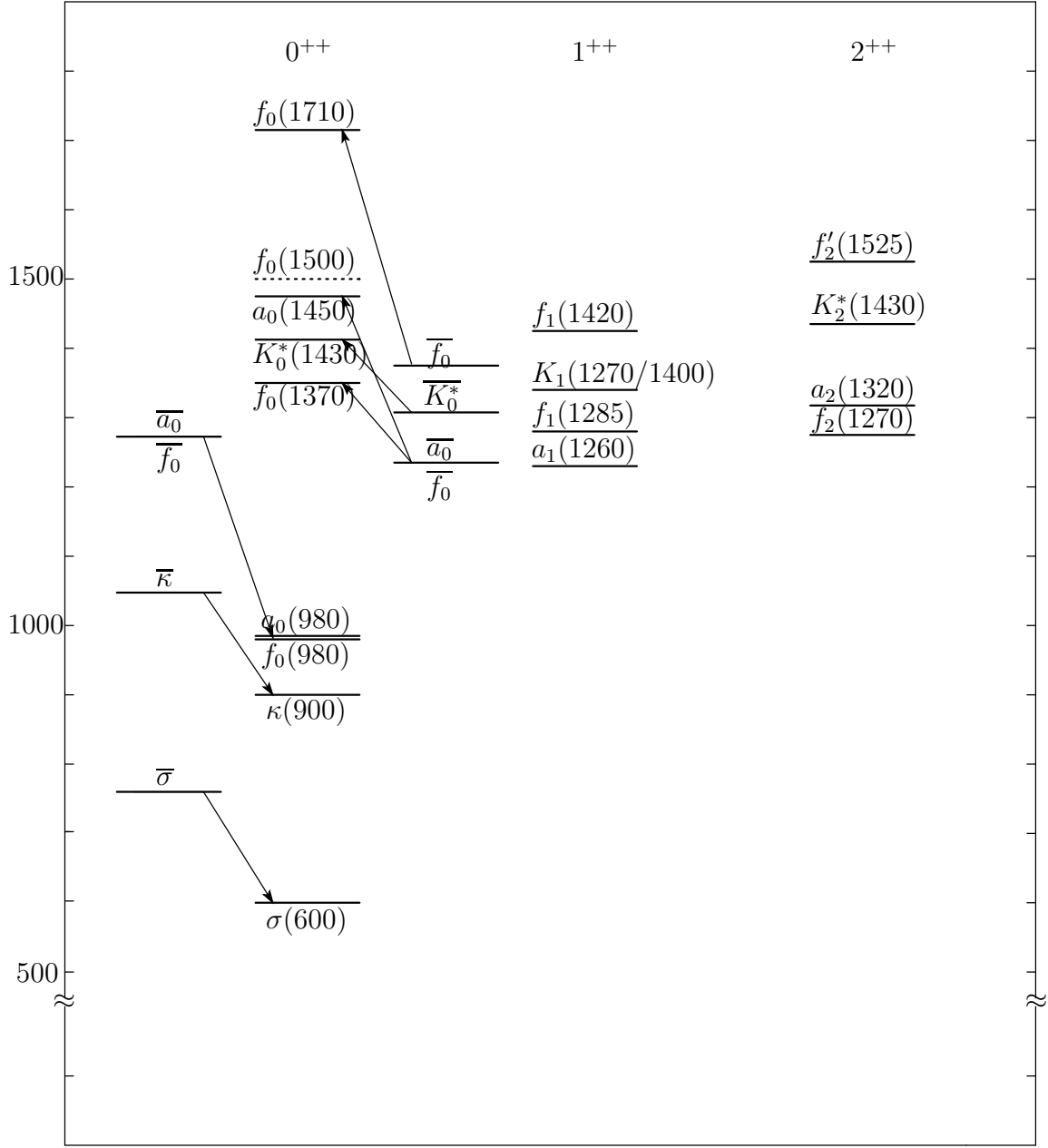}
\end{picture}
\vspace{17.0cm}
\caption{The $2^{++}$, $1^{++}$ and $0^{++}$ meson mass spectra. The particles 
with overline are particles before-mixing with masses estimated using the 
masses of $0^{++}$, $1^{++}$ and $2^{++}$ mesons. (See the text).}
\end{figure}
If we express the transition strength between $\overline{a_0(980)}$ 
and $\overline{a_0(1450)}$ as $\lambda_{a_0}$, then the mass matrix is written 
as
\begin{equation}
\left(\begin{array}{cc}
m^2_{\overline{a_0(980)}}&\lambda_{a_0}\\
\lambda_{a_0}&m^2_{\overline{a_0(1450)}}
\end{array}\right),\ \ m_{\overline{a_0(980)}}=1.271{\rm GeV},\ \ 
m_{\overline{a_0(1450)}}=1.236{\rm GeV}
\end{equation}
and this has the eigenvalues $m_{a_0(980)}=0.985{\rm GeV}$ and 
$m_{a_0(1450)}=1.474{\rm GeV}$ at the $\lambda_{a_0}=0.600
{\rm GeV}^2$. Mixing angle is evaluated as $\theta_{a_0}=47.1^{\circ}$. 
Similarly, for $I=1/2$ scalar mesons, mixing matrix 
\begin{equation}
\left(\begin{array}{cc}
m^2_{\overline{\kappa(900)}}&\lambda_{K_0}\\
\lambda_{K_0}&m^2_{\overline{K_0^*(1430)}}
\end{array}\right),\ \ m_{\overline{\kappa(900)}}=1.047{\rm GeV},\ \ 
m_{\overline{K_0^*(1430)}}=1.307{\rm GeV}
\end{equation}
has eigenvalues $m_{\kappa(900)}=0.900{\rm GeV}$ and $m_{K^*_0(1430)}=
1.412{\rm GeV}$ at the $\lambda_{K_0}=0.507{\rm GeV}^2$ and mixing angle 
$\theta_{K_0}=29.5^\circ$. 
\par 
The obtained transition strength has two characteristics: (1) The strength is 
very large and (2) the values $\lambda_{a_0}$ for $I=1$ and $\lambda_{K_0}$ 
for $I=1/2$ are nearly equal. These characteristics are explained easily by 
assuming the structure of light scalar mesons which is explained explicitly in 
next section.
\section{Structure of light scalar mesons}
Because the scalar mesons are considered as an $SU(3)$ nonet, we present 
the scalar meson field as $N'^b_a$ for conventional $L=1\ q\bar{q}$ scalar 
mesons and $N^b_a$ for light scalar mesons. In representation $N'^b_a$, 
$a$ and $b$ denote the $SU(3)$ indices of triplet quark field $q_a$ and 
unti-triplet quark field $\bar{q}^b$ respectively. 
\begin{equation}
N'^b_a\sim q_a\bar{q}^b\ \  {\rm for}\ L=1\ q\bar{q}\ 
              {\rm scalar\ mesons.}
\end{equation}
For $N^b_a$, $a$ and $b$ are considered as $SU(3)$ indices of $q_a$ and 
$\bar{q}^b$ when the light scalar mesons are considered as chiral scalar 
partner of pseudoscalar nonet with $q\bar{q}$ structure. In the case where 
the light scalar mesons are considered as $qq\bar{q}\bar{q}$, $a$ and $b$ in 
representation $N^b_a$ denote the $SU(3)$ indices of "dual" quark $T_a=
\epsilon_{abc}\bar{q}^b\bar{q}^c$ and "dual" anti-quark $\bar{T}^a=
\epsilon^{abc}q_bq_c$, respectively\cite{JAFFE,SCHECHTER}.
\begin{eqnarray}
&&N^a_b\sim q_b\bar{q}^a\ \ {\rm for}\ q\bar{q}\ {\rm light\ scalar\ mesons,}\\
&&N^a_b\sim T_b\bar{T}^a\sim \epsilon_{bde}\bar{q}^d\bar{q}^e
\epsilon^{abc}q_bq_c\ \ {\rm for}\ qq\bar{q}\bar{q}\ 
{\rm light\ scalar\ mesons.}
\end{eqnarray}
Here, we show the explicit flavor configuration of the $qq\bar{q}\bar{q}$ 
scalar mesons to see what content of quarks are included in each scalar 
mesons, 
\begin{equation}
\begin{array}{ccccc}
    a_0^+&\ \sim\ &N_1^2&\ \sim\ &\bar{s}\bar{d}us\\
    a_0^0&\ \sim\ &\frac{1}{\sqrt{2}}(N_1^1-N_2^2)&\ \sim\ &
    \frac{1}{\sqrt{2}}(\bar{s}\bar{d}ds-\bar{s}\bar{u}us)\\
    a_0^-&\ \sim\ &N_2^1&\ \sim\ &\bar{s}\bar{u}ds\\
    \kappa^+&\ \sim\ &N_1^3&\ \sim\ &\bar{s}\bar{d}ud\\
    \kappa^0&\ \sim\ &N_2^3&\ \sim\ &\bar{s}\bar{u}ud\\
    \overline{\kappa}^0&\ \sim\ &N_3^2&\ \sim\ &\bar{u}\bar{d}us\\
    \kappa^-&\ \sim\ &N_3^1&\ \sim\ &\bar{u}\bar{d}ds\\
    f_0&\ \sim\ &\frac{1}{\sqrt{2}}(N_1^1+N_2^2)&\ \sim\ &
    \frac{1}{\sqrt{2}}(\bar{s}\bar{d}ds+\bar{s}\bar{u}us)\\
    \sigma&\ \sim\ &N_3^3&\ \sim\ &\bar{u}\bar{d}ud
    \end{array}
\end{equation}
in the ideal mixing limit. Here, we assume that the state $f_0$ is 
$\frac{1}{\sqrt{2}}(N_1^1+N_2^2)\sim\frac{1}{\sqrt{2}}(\bar{s}\bar{d}ds+
\bar{s}\bar{u}us)$ and $\sigma$ is $N_3^3\sim\bar{u}\bar{d}ud$ from the 
following consideration of masses of light scalar mesons. 
\par
We assume that the masses of light scalar mesons are described by the 
following chiral symmetric effective Lagrangian density 
\begin{equation}
L^{eff}=-a{\rm Tr}(NN)-b{\rm Tr}(NNM)-\frac12\lambda{\rm Tr}(N)
{\rm Tr}(N),
\end{equation}
where $M$ is the "spurion matrix" $M={\rm diag}[1,1,x]$, $x$ representing the 
symmetry breaking of the strange quark mass to non-strange quark mass. 
This formula is adopted in Black {\it et al} 's analysis \cite{SCHECHTER}. 
This Lagrangian is equivalent to the generalized mass matrix used in our 
previous analysis of $q\bar{q}$ scalar mesons \cite{TESHIMA}. If light scalar 
mesons are $q\bar{q}$ mesons, the third term $\lambda{\rm Tr}(N){\rm Tr}(N)$ 
denotes the transition amplitudes reproducing the $U(1)$ anomaly term in 
pseudoscalar meson case corresponding to the violation term of OZI rule in 2nd 
order shown in the Fig.~2. 
\begin{figure}[htb]
\vspace*{2.5cm}
\begin{picture}(10,5)
\hspace*{4cm}
\unitlength 0.1in
\begin{picture}( 26.0000, 11.0300)(  3.0000,-12.3300)
\special{pn 13}%
\special{pa 1416 414}%
\special{pa 1420 410}%
\special{pa 1426 406}%
\special{pa 1430 400}%
\special{pa 1436 396}%
\special{pa 1440 394}%
\special{pa 1446 390}%
\special{pa 1450 388}%
\special{pa 1456 388}%
\special{pa 1460 386}%
\special{pa 1466 386}%
\special{pa 1470 388}%
\special{pa 1476 390}%
\special{pa 1480 392}%
\special{pa 1486 396}%
\special{pa 1490 400}%
\special{pa 1496 404}%
\special{pa 1500 408}%
\special{pa 1506 412}%
\special{pa 1510 416}%
\special{pa 1516 422}%
\special{pa 1520 424}%
\special{pa 1526 428}%
\special{pa 1530 430}%
\special{pa 1536 432}%
\special{pa 1540 434}%
\special{pa 1546 434}%
\special{pa 1550 434}%
\special{pa 1556 432}%
\special{pa 1560 430}%
\special{pa 1566 428}%
\special{pa 1570 424}%
\special{pa 1576 420}%
\special{pa 1580 416}%
\special{pa 1586 412}%
\special{pa 1590 408}%
\special{pa 1596 404}%
\special{pa 1600 400}%
\special{pa 1606 396}%
\special{pa 1610 392}%
\special{pa 1616 390}%
\special{pa 1620 388}%
\special{pa 1626 386}%
\special{pa 1630 386}%
\special{pa 1636 388}%
\special{pa 1640 388}%
\special{pa 1646 390}%
\special{pa 1650 394}%
\special{pa 1656 398}%
\special{pa 1660 402}%
\special{pa 1666 406}%
\special{pa 1670 410}%
\special{pa 1676 414}%
\special{pa 1680 418}%
\special{pa 1686 422}%
\special{pa 1690 426}%
\special{pa 1696 430}%
\special{pa 1700 432}%
\special{pa 1706 434}%
\special{pa 1710 434}%
\special{pa 1716 434}%
\special{pa 1720 434}%
\special{pa 1726 432}%
\special{pa 1730 430}%
\special{pa 1736 426}%
\special{pa 1740 422}%
\special{pa 1746 418}%
\special{pa 1750 414}%
\special{pa 1756 410}%
\special{pa 1760 406}%
\special{pa 1766 402}%
\special{pa 1770 398}%
\special{pa 1776 394}%
\special{pa 1780 390}%
\special{pa 1786 388}%
\special{pa 1790 388}%
\special{pa 1796 386}%
\special{pa 1800 386}%
\special{pa 1806 388}%
\special{pa 1810 390}%
\special{pa 1816 392}%
\special{pa 1820 396}%
\special{pa 1826 398}%
\special{pa 1830 404}%
\special{pa 1836 408}%
\special{pa 1840 412}%
\special{pa 1846 416}%
\special{pa 1850 420}%
\special{pa 1856 424}%
\special{pa 1860 428}%
\special{pa 1866 430}%
\special{pa 1870 432}%
\special{pa 1876 434}%
\special{pa 1880 434}%
\special{pa 1886 434}%
\special{pa 1890 432}%
\special{pa 1896 430}%
\special{pa 1900 428}%
\special{pa 1906 424}%
\special{pa 1910 422}%
\special{pa 1916 416}%
\special{pa 1920 412}%
\special{pa 1926 408}%
\special{pa 1930 404}%
\special{pa 1936 400}%
\special{pa 1940 396}%
\special{pa 1946 392}%
\special{pa 1950 390}%
\special{pa 1956 388}%
\special{sp}%
%
\special{pn 13}%
\special{pa 620 396}%
\special{pa 1426 396}%
\special{fp}%
\special{pa 1426 996}%
\special{pa 1426 996}%
\special{fp}%
\special{pa 620 996}%
\special{pa 620 996}%
\special{fp}%
\special{pa 620 996}%
\special{pa 1412 996}%
\special{fp}%
\special{pa 1412 996}%
\special{pa 1412 396}%
\special{fp}%
\special{pa 1948 396}%
\special{pa 2754 396}%
\special{fp}%
\special{pa 2754 996}%
\special{pa 1948 996}%
\special{fp}%
\special{pa 1948 996}%
\special{pa 1948 396}%
\special{fp}%
\special{pn 13}%
\special{pa 1406 1008}%
\special{pa 1410 1004}%
\special{pa 1416 1000}%
\special{pa 1420 996}%
\special{pa 1426 992}%
\special{pa 1430 988}%
\special{pa 1436 984}%
\special{pa 1440 980}%
\special{pa 1446 976}%
\special{pa 1450 974}%
\special{pa 1456 972}%
\special{pa 1460 970}%
\special{pa 1466 970}%
\special{pa 1470 972}%
\special{pa 1476 972}%
\special{pa 1480 974}%
\special{pa 1486 978}%
\special{pa 1490 980}%
\special{pa 1496 984}%
\special{pa 1500 990}%
\special{pa 1506 994}%
\special{pa 1510 998}%
\special{pa 1516 1002}%
\special{pa 1520 1006}%
\special{pa 1526 1010}%
\special{pa 1530 1014}%
\special{pa 1536 1016}%
\special{pa 1540 1018}%
\special{pa 1546 1018}%
\special{pa 1550 1018}%
\special{pa 1556 1018}%
\special{pa 1560 1016}%
\special{pa 1566 1014}%
\special{pa 1570 1010}%
\special{pa 1576 1008}%
\special{pa 1580 1004}%
\special{pa 1586 998}%
\special{pa 1590 994}%
\special{pa 1596 990}%
\special{pa 1600 986}%
\special{pa 1606 982}%
\special{pa 1610 978}%
\special{pa 1616 974}%
\special{pa 1620 972}%
\special{pa 1626 972}%
\special{pa 1630 970}%
\special{pa 1636 970}%
\special{pa 1640 972}%
\special{pa 1646 974}%
\special{pa 1650 976}%
\special{pa 1656 980}%
\special{pa 1660 982}%
\special{pa 1666 986}%
\special{pa 1670 992}%
\special{pa 1676 996}%
\special{pa 1680 1000}%
\special{pa 1686 1004}%
\special{pa 1690 1008}%
\special{pa 1696 1012}%
\special{pa 1700 1014}%
\special{pa 1706 1016}%
\special{pa 1710 1018}%
\special{pa 1716 1018}%
\special{pa 1720 1018}%
\special{pa 1726 1018}%
\special{pa 1730 1016}%
\special{pa 1736 1012}%
\special{pa 1740 1010}%
\special{pa 1746 1006}%
\special{pa 1750 1002}%
\special{pa 1756 996}%
\special{pa 1760 992}%
\special{pa 1766 988}%
\special{pa 1770 984}%
\special{pa 1776 980}%
\special{pa 1780 976}%
\special{pa 1786 974}%
\special{pa 1790 972}%
\special{pa 1796 970}%
\special{pa 1800 970}%
\special{pa 1806 970}%
\special{pa 1810 972}%
\special{pa 1816 974}%
\special{pa 1820 978}%
\special{pa 1826 980}%
\special{pa 1830 984}%
\special{pa 1836 988}%
\special{pa 1840 994}%
\special{pa 1846 998}%
\special{pa 1850 1002}%
\special{pa 1856 1006}%
\special{pa 1860 1010}%
\special{pa 1866 1014}%
\special{pa 1870 1016}%
\special{pa 1876 1018}%
\special{pa 1880 1018}%
\special{pa 1886 1018}%
\special{pa 1890 1018}%
\special{pa 1896 1016}%
\special{pa 1900 1014}%
\special{pa 1906 1012}%
\special{pa 1910 1008}%
\special{pa 1916 1004}%
\special{pa 1920 1000}%
\special{pa 1926 994}%
\special{pa 1930 990}%
\special{pa 1936 986}%
\special{pa 1940 982}%
\special{sp}%
%
\special{pn 8}%
\special{pa 1136 410}%
\special{pa 600 410}%
\special{pa 600 1012}%
\special{pa 1136 1012}%
\special{pa 1136 410}%
\special{ip}%
%
\special{pn 8}%
\special{pa 2224 410}%
\special{pa 2760 410}%
\special{pa 2760 1012}%
\special{pa 2224 1012}%
\special{pa 2224 410}%
\special{ip}%
%
\special{pn 8}%
\special{pa 2760 458}%
\special{pa 2224 940}%
\special{fp}%
\special{pa 2760 604}%
\special{pa 2316 998}%
\special{fp}%
\special{pa 2760 746}%
\special{pa 2478 998}%
\special{fp}%
\special{pa 2760 890}%
\special{pa 2640 998}%
\special{fp}%
\special{pa 2654 410}%
\special{pa 2224 796}%
\special{fp}%
\special{pa 2490 410}%
\special{pa 2224 650}%
\special{fp}%
\special{pa 2332 410}%
\special{pa 2224 506}%
\special{fp}%
%
\special{pn 4}%
\special{pa 1136 614}%
\special{pa 710 998}%
\special{fp}%
\special{pa 1136 472}%
\special{pa 600 950}%
\special{fp}%
\special{pa 1044 410}%
\special{pa 600 806}%
\special{fp}%
\special{pa 882 410}%
\special{pa 600 662}%
\special{fp}%
\special{pa 722 410}%
\special{pa 600 518}%
\special{fp}%
\special{pa 1136 758}%
\special{pa 868 998}%
\special{fp}%
\special{pa 1136 904}%
\special{pa 1030 998}%
\special{fp}%
%
\special{pn 8}%
\special{pa 2236 410}%
\special{pa 2236 1012}%
\special{ip}%
%
\special{pn 8}%
\special{pa 2214 394}%
\special{pa 2214 996}%
\special{fp}%
\special{pa 1140 996}%
\special{pa 1140 394}%
\special{fp}%
\put(3.0000,-8.0000){\makebox(0,0)[lb]{$q\bar{q}$}}%
\put(29.0000,-8.0000){\makebox(0,0)[lb]{$q\bar{q}$}}%
\put(16.0000,-3.0000){\makebox(0,0)[lb]{$g$}}%
\put(16.0000,-12.5000){\makebox(0,0)[lb]{$g$}}%
%
\special{pn 13}%
\special{pa 990 400}%
\special{pa 1000 400}%
\special{fp}%
\special{sh 1}%
\special{pa 1000 400}%
\special{pa 934 380}%
\special{pa 948 400}%
\special{pa 934 420}%
\special{pa 1000 400}%
\special{fp}%
%
\special{pn 13}%
\special{pa 930 1000}%
\special{pa 920 1000}%
\special{fp}%
\special{sh 1}%
\special{pa 920 1000}%
\special{pa 988 1020}%
\special{pa 974 1000}%
\special{pa 988 980}%
\special{pa 920 1000}%
\special{fp}%
%
\special{pn 13}%
\special{pa 2390 400}%
\special{pa 2400 400}%
\special{fp}%
\special{sh 1}%
\special{pa 2400 400}%
\special{pa 2334 380}%
\special{pa 2348 400}%
\special{pa 2334 420}%
\special{pa 2400 400}%
\special{fp}%
%
\special{pn 13}%
\special{pa 2330 1000}%
\special{pa 2320 1000}%
\special{fp}%
\special{sh 1}%
\special{pa 2320 1000}%
\special{pa 2388 1020}%
\special{pa 2374 1000}%
\special{pa 2388 980}%
\special{pa 2320 1000}%
\special{fp}%
\end{picture}%
\end{picture}
\caption{Graph for the $q\bar{q}$ meson transition violating the 
OZI rule in 2nd order. Straight lines represent the quarks and wavy lines the 
gluons. Oblique lines represent the gluon interactions in $q\bar{q}$ bound 
state. }
\end{figure}
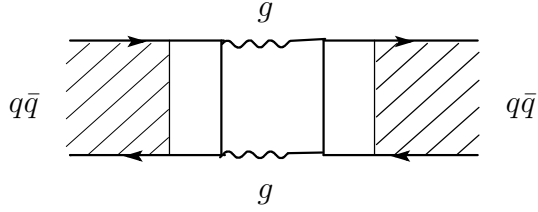
\par 
We review the analysis for scalar meson masses using the effective Lagrangian 
(12) disregarding the inter-mixing. First, we assume that the $\sigma(600)$ is 
nearly equal to ideal state $\frac{N^1_1+N^2_2}{\sqrt{2}}$ and the $f_0(980)$ 
to ideal state $N^3_3$, because the $f_0(980)$ has the strong $s\bar{s}$ like 
structure. By using the Eq.~(12), the scalar meson masses are represented as 
\begin{eqnarray}
&&m_{a}^2=2a+2b,\ m_{\kappa}^2=2a+(1+x)b,\\
&&\left(\begin{array}{c}
    \sigma(600)\\
    f_0(980)
    \end{array}\right)=
    O
    \left(\begin{array}{c}
    \frac{N^1_1+N^2_2}{\sqrt{2}}\\
    N^3_3
    \end{array}\right)\sim
    \left(\begin{array}{c}
    \frac{N^1_1+N^2_2}{\sqrt{2}}\\
    N^3_3
    \end{array}\right),\nonumber\\
&&\hspace{1cm}O=\left(\begin{array}{cc}
    \cos\theta&-\sin\theta\\
    \sin\theta&\cos\theta
    \end{array}\right),\nonumber\\
&&O\left(\begin{array}{cc}
    m_a^2+2\lambda&\sqrt{2}\lambda\\
    \sqrt{2}\lambda&2m^2_{\kappa}-m^2_a+\lambda
    \end{array}\right){^tO}=
    \left(\begin{array}{cc}
    m_{\sigma}^2&0\\
    0&m^2_{f_0}
    \end{array}\right).
 \end{eqnarray}   
From the relation $2m^2_{\kappa}+3\lambda=m^2_{\sigma}+m^2_{f_0}$, we obtain 
$\lambda=-0.0999~{\rm GeV^2}$ and eigenvalues 694MeV and 916MeV. 
One eigenvalues 694MeV of these is not so close to the values 600MeV, 
furthermore mixing angle is $64^\circ$ and $f_0$ state is far from the ideal 
$s\bar{s}$ state. In Black {\it et al.}'s analysis , the $SU(3)$ 
breaking correction is introduced 
by adding the term ${\rm Tr}(NM){\rm Tr}(NM)$ in Eq.~(12) \cite{SCHECHTER}. 
In our analysis \cite{TESHIMA}, the $SU(3)$ breaking correction is introduced 
by multiplying $k$ to $\lambda$ 
as  
$$
O\left(\begin{array}{cc}
    m_a^2+2\lambda&k\sqrt{2}\lambda\\
    k\sqrt{2}\lambda&2m^2_{\kappa}-m^2_a+k^2\lambda
    \end{array}\right){^tO}=
    \left(\begin{array}{cc}
    m_{\sigma}^2&0\\
    0&m^2_{f_0}
    \end{array}\right).
$$ 
In this case, if $\lambda=-0.00935~{\rm GeV^2}$ and $k=5.48$, we can 
obtain the eigenvalues exactly equal to the values $m_{\sigma}=600$MeV and 
$m_{f_0}=980$MeV, but mixing angle is $83^\circ$ and $f_0$ state is almost 
non-$s\bar{s}$ state. Thus, it is difficult to consider that the light scalar 
mesons are $q\bar{q}$ states. 
\par
Next we consider the case that the light scalar mesons are $qq\bar{q}\bar{q}$.
In this case, the third term $\frac12\lambda{\rm Tr}(N){\rm Tr}(N)$ represents 
also the OZI violation terms in 2nd order as shown in Fig.~3.
\begin{figure}[htb]
\vspace*{2.5cm}
\begin{picture}(10,4)
\hspace*{4cm}
\unitlength 0.1in
\begin{picture}( 25.0000,  6.1000)(  2.0000,-10.1000)
%
\special{pn 13}%
\special{pa 600 410}%
\special{pa 2600 410}%
\special{fp}%
\special{pa 2600 1010}%
\special{pa 600 1010}%
\special{fp}%
\special{pa 600 610}%
\special{pa 1300 610}%
\special{fp}%
\special{pa 1300 810}%
\special{pa 600 810}%
\special{fp}%
\special{pa 2600 610}%
\special{pa 1900 610}%
\special{fp}%
\special{pa 1900 810}%
\special{pa 2600 810}%
\special{fp}%
%
\special{pn 13}%
\special{ar 1900 710 100 100  1.5707963 4.7123890}%
%
\special{pn 13}%
\special{ar 1300 710 100 100  4.7123890 6.2831853}%
\special{ar 1300 710 100 100  0.0000000 1.5707963}%
\special{pn 13}%
\special{pa 1400 700}%
\special{pa 1406 704}%
\special{pa 1410 708}%
\special{pa 1416 714}%
\special{pa 1420 718}%
\special{pa 1426 722}%
\special{pa 1430 726}%
\special{pa 1436 728}%
\special{pa 1440 730}%
\special{pa 1446 730}%
\special{pa 1450 730}%
\special{pa 1456 726}%
\special{pa 1460 724}%
\special{pa 1466 720}%
\special{pa 1470 714}%
\special{pa 1476 710}%
\special{pa 1480 704}%
\special{pa 1486 700}%
\special{pa 1490 696}%
\special{pa 1496 694}%
\special{pa 1500 692}%
\special{pa 1506 690}%
\special{pa 1510 690}%
\special{pa 1516 692}%
\special{pa 1520 696}%
\special{pa 1526 700}%
\special{pa 1530 704}%
\special{pa 1536 708}%
\special{pa 1540 714}%
\special{pa 1546 718}%
\special{pa 1550 722}%
\special{pa 1556 726}%
\special{pa 1560 728}%
\special{pa 1566 730}%
\special{pa 1570 730}%
\special{pa 1576 730}%
\special{pa 1580 728}%
\special{pa 1586 724}%
\special{pa 1590 720}%
\special{pa 1596 716}%
\special{pa 1600 710}%
\special{pa 1606 706}%
\special{pa 1610 700}%
\special{pa 1616 696}%
\special{pa 1620 694}%
\special{pa 1626 692}%
\special{pa 1630 690}%
\special{pa 1636 690}%
\special{pa 1640 692}%
\special{pa 1646 694}%
\special{pa 1650 698}%
\special{pa 1656 702}%
\special{pa 1660 708}%
\special{pa 1666 712}%
\special{pa 1670 718}%
\special{pa 1676 722}%
\special{pa 1680 726}%
\special{pa 1686 728}%
\special{pa 1690 730}%
\special{pa 1696 730}%
\special{pa 1700 730}%
\special{pa 1706 728}%
\special{pa 1710 724}%
\special{pa 1716 720}%
\special{pa 1720 716}%
\special{pa 1726 712}%
\special{pa 1730 706}%
\special{pa 1736 702}%
\special{pa 1740 698}%
\special{pa 1746 694}%
\special{pa 1750 692}%
\special{pa 1756 690}%
\special{pa 1760 690}%
\special{pa 1766 692}%
\special{pa 1770 694}%
\special{pa 1776 698}%
\special{pa 1780 702}%
\special{pa 1786 708}%
\special{pa 1790 712}%
\special{pa 1796 716}%
\special{pa 1800 722}%
\special{sp}%
%
\special{pn 13}%
\special{pa 600 410}%
\special{pa 1200 410}%
\special{pa 1200 1010}%
\special{pa 600 1010}%
\special{pa 600 410}%
\special{ip}%
%
\special{pn 13}%
\special{pa 590 410}%
\special{pa 1190 410}%
\special{pa 1190 1010}%
\special{pa 590 1010}%
\special{pa 590 410}%
\special{ip}%
%
\special{pn 4}%
\special{pa 1120 810}%
\special{pa 930 1000}%
\special{fp}%
\special{pa 1000 810}%
\special{pa 810 1000}%
\special{fp}%
\special{pa 880 810}%
\special{pa 690 1000}%
\special{fp}%
\special{pa 760 810}%
\special{pa 600 970}%
\special{fp}%
\special{pa 640 810}%
\special{pa 600 850}%
\special{fp}%
\special{pa 1180 870}%
\special{pa 1050 1000}%
\special{fp}%
%
\special{pn 4}%
\special{pa 1080 610}%
\special{pa 890 800}%
\special{fp}%
\special{pa 960 610}%
\special{pa 770 800}%
\special{fp}%
\special{pa 840 610}%
\special{pa 650 800}%
\special{fp}%
\special{pa 720 610}%
\special{pa 600 730}%
\special{fp}%
\special{pa 1180 630}%
\special{pa 1010 800}%
\special{fp}%
\special{pa 1180 750}%
\special{pa 1130 800}%
\special{fp}%
%
\special{pn 4}%
\special{pa 1040 410}%
\special{pa 850 600}%
\special{fp}%
\special{pa 920 410}%
\special{pa 730 600}%
\special{fp}%
\special{pa 800 410}%
\special{pa 610 600}%
\special{fp}%
\special{pa 680 410}%
\special{pa 600 490}%
\special{fp}%
\special{pa 1160 410}%
\special{pa 970 600}%
\special{fp}%
\special{pa 1180 510}%
\special{pa 1090 600}%
\special{fp}%
%
\special{pn 8}%
\special{pa 1200 410}%
\special{pa 1200 1010}%
\special{fp}%
\special{pa 2000 1010}%
\special{pa 2000 410}%
\special{fp}%
%
\special{pn 8}%
\special{pa 2000 410}%
\special{pa 2600 410}%
\special{pa 2600 1010}%
\special{pa 2000 1010}%
\special{pa 2000 410}%
\special{ip}%
%
\special{pn 8}%
\special{pa 2000 410}%
\special{pa 2600 410}%
\special{pa 2600 1010}%
\special{pa 2000 1010}%
\special{pa 2000 410}%
\special{ip}%
%
\special{pn 4}%
\special{pa 2440 810}%
\special{pa 2250 1000}%
\special{fp}%
\special{pa 2320 810}%
\special{pa 2130 1000}%
\special{fp}%
\special{pa 2200 810}%
\special{pa 2010 1000}%
\special{fp}%
\special{pa 2080 810}%
\special{pa 2000 890}%
\special{fp}%
\special{pa 2560 810}%
\special{pa 2370 1000}%
\special{fp}%
\special{pa 2600 890}%
\special{pa 2490 1000}%
\special{fp}%
%
\special{pn 8}%
\special{pa 2000 400}%
\special{pa 2600 400}%
\special{pa 2600 800}%
\special{pa 2000 800}%
\special{pa 2000 400}%
\special{ip}%
%
\special{pn 4}%
\special{pa 2280 610}%
\special{pa 2090 800}%
\special{fp}%
\special{pa 2160 610}%
\special{pa 2000 770}%
\special{fp}%
\special{pa 2040 610}%
\special{pa 2000 650}%
\special{fp}%
\special{pa 2400 610}%
\special{pa 2210 800}%
\special{fp}%
\special{pa 2520 610}%
\special{pa 2330 800}%
\special{fp}%
\special{pa 2600 650}%
\special{pa 2450 800}%
\special{fp}%
%
\special{pn 4}%
\special{pa 2360 410}%
\special{pa 2170 600}%
\special{fp}%
\special{pa 2240 410}%
\special{pa 2050 600}%
\special{fp}%
\special{pa 2120 410}%
\special{pa 2000 530}%
\special{fp}%
\special{pa 2480 410}%
\special{pa 2290 600}%
\special{fp}%
\special{pa 2590 420}%
\special{pa 2410 600}%
\special{fp}%
\special{pa 2600 530}%
\special{pa 2530 600}%
\special{fp}%
%
\special{pn 13}%
\special{pa 1200 610}%
\special{pa 1300 610}%
\special{fp}%
\special{sh 1}%
\special{pa 1300 610}%
\special{pa 1234 590}%
\special{pa 1248 610}%
\special{pa 1234 630}%
\special{pa 1300 610}%
\special{fp}%
\special{pa 1300 810}%
\special{pa 1200 810}%
\special{fp}%
\special{sh 1}%
\special{pa 1200 810}%
\special{pa 1268 830}%
\special{pa 1254 810}%
\special{pa 1268 790}%
\special{pa 1200 810}%
\special{fp}%
\special{pa 1880 610}%
\special{pa 1980 610}%
\special{fp}%
\special{sh 1}%
\special{pa 1980 610}%
\special{pa 1914 590}%
\special{pa 1928 610}%
\special{pa 1914 630}%
\special{pa 1980 610}%
\special{fp}%
\special{pa 1980 810}%
\special{pa 1890 810}%
\special{fp}%
\special{sh 1}%
\special{pa 1890 810}%
\special{pa 1958 830}%
\special{pa 1944 810}%
\special{pa 1958 790}%
\special{pa 1890 810}%
\special{fp}%
\put(27.0000,-7.6000){\makebox(0,0)[lb]{$qq\bar{q}\bar{q}$}}%
\put(2.0000,-7.6000){\makebox(0,0)[lb]{$qq\bar{q}\bar{q}$}}%
%
\special{pn 13}%
\special{pa 1800 1010}%
\special{pa 1600 1010}%
\special{fp}%
\special{sh 1}%
\special{pa 1600 1010}%
\special{pa 1668 1030}%
\special{pa 1654 1010}%
\special{pa 1668 990}%
\special{pa 1600 1010}%
\special{fp}%
\special{pa 1400 410}%
\special{pa 1600 410}%
\special{fp}%
\special{sh 1}%
\special{pa 1600 410}%
\special{pa 1534 390}%
\special{pa 1548 410}%
\special{pa 1534 430}%
\special{pa 1600 410}%
\special{fp}%
\end{picture}%
\end{picture}
\caption{Graph for the $qq\bar{q}\bar{q}$ meson transition violating 
the OZI rule in 2nd order.}
\end{figure}
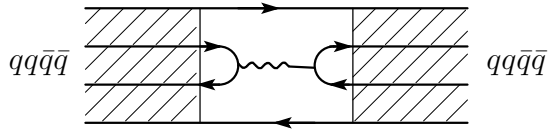
In this case, we adopt the quark configuration in Eq.~(11) because the $f_0$ 
state is almost $s\bar{s}$ like. The masses of this case are represented 
in the same form as the $q\bar{q}$ case except for the configuration of 
$f_0(980)$ and $\sigma(600)$ which are nearly 
$f_N\equiv\frac{N^1_1+N^2_2}{\sqrt{2}}\sim\frac{1}{\sqrt{2}}(\bar{s}\bar{d}ds+
\bar{s}\bar{u}us)$ and $f_S\equiv N^3_3\sim\bar{u}\bar{d}ud$, respectively;
\begin{eqnarray}
&&m_{a}^2=2a+2b,\ m_{\kappa}^2=2a+(1+x)b,\\
&&\left(\begin{array}{c}
    f_0(980)\\
    \sigma(600)
    \end{array}\right)=
    O
    \left(\begin{array}{c}
    \frac{N^1_1+N^2_2}{\sqrt{2}}\\
    N^3_3
    \end{array}\right)\sim
    \left(\begin{array}{c}
    \frac{N^1_1+N^2_2}{\sqrt{2}}\\
    N^3_3
    \end{array}\right),\nonumber\\
&&\hspace{1cm}O=\left(\begin{array}{cc}
    \cos\theta&-\sin\theta\\
    \sin\theta&\cos\theta
    \end{array}\right),\nonumber\\
&&O\left(\begin{array}{cc}
    m_a^2+2\lambda&\sqrt{2}\lambda\\
    \sqrt{2}\lambda&2m^2_{\kappa}-m^2_a+\lambda
    \end{array}\right){^tO}=
    \left(\begin{array}{cc}
    m^2_{f_0}&0\\
    0&m_{\sigma}^2
    \end{array}\right).
\end{eqnarray}   
Here, it should be noted that the first and second terms in effective 
Lagrangian (12) contain the contributions corresponding to the transition 
represented by the graph in Fig~(3). In fact, the masses for $a^+_0$, $\kappa^+$,
$f_N$, $f_S$ mesons and transition amplitude from $f_N$ to $f_S$ 
are represented as
\begin{eqnarray}
&&m_{a_0^+}^2=<\bar{s}\bar{d}us|H|\bar{s}\bar{d}us>=2a+2b=M_0+2(m_u+m_s)+
<\bar{u}u|\bar{u}u>\lambda_{ss},\nonumber\\
&&m_{\kappa^+}^2=<\bar{s}\bar{d}ud|H|\bar{s}\bar{d}ud>=2a+(1+x)b=M_0+3m_u+m_s+
<\bar{s}u|\bar{s}u>\lambda_{uu},\nonumber\\
&&m_{f_N}^2=<\frac1{\sqrt{2}}(\bar{s}\bar{d}ds+\bar{s}\bar{u}us)|H|
\frac1{\sqrt{2}}(\bar{s}\bar{d}ds+\bar{s}\bar{u}us)>=2a+2b+2\lambda\nonumber\\
&&\hspace{3cm}=M_0+2(m_u+m_s)+<\bar{u}u|\bar{u}u>\lambda_{ss}+
2<\bar{s}s|\bar{s}s>\lambda_{uu},\nonumber\\
&&m_{f_S}^2=<\bar{u}\bar{d}ud|H|\bar{u}\bar{d}ud>=2a+2bx+\lambda=M_0+4m_u+
2<\bar{u}u|\bar{u}u>\lambda_{uu},\nonumber\\
&&\sqrt{2}\lambda=<\frac1{\sqrt{2}}(\bar{s}\bar{d}ds
+\bar{s}\bar{u}us)|H|\bar{u}\bar{d}ud>=\sqrt{2}<\bar{u}u|\bar{u}u>
\lambda_{su},
\end{eqnarray}
where $M_0$, $m_u$ and $m_s$ are the terms proportional to the potential 
energy, $u(d)$ quark mass and $s$ quark mass, respectively and last term 
represents the contribution of the OZI violation term in 2nd order. In 
obtaining Eq.~(16), we assumed that the $SU(3)$ symmetry breaking is 
negligible for the contribution of the OZI violation term and then 
$<\bar{u}u|\bar{u}u>\lambda_{uu}=<\bar{s}u|\bar{s}u>\lambda_{uu}
=<\bar{s}s|\bar{s}s>\lambda_{uu}=<\bar{u}u|\bar{u}u>\lambda_{ss}$ were assumed. \par
Diagonalising the mass matrix (16), we obtain the mixing angle as 
$26^\circ$ and $\lambda=-0.0999~{\rm GeV^2}$ and $f_0$ nearly equal to 
$f_N=\frac1{\sqrt{2}}(\bar{s}\bar{d}ds+\bar{s}\bar{u}us)$ is obtained. 
To obtain the masses of $f_0(980)$ and $\sigma(600)$ exactly equal to 
980 MeV and 600 MeV, we consider the inter-mixing in next section.
From the analysis for the masses of light scalar mesons, we can conclude 
that the light scalar mesons prefer the $qq\bar{q}\bar{q}$ 
structure to $q\bar{q}$ one. Lastly, it should be noted that the strength of 
the transition amplitude $|\lambda|=0.0999~{\rm GeV^2}$ is rather small 
compared with that of the inter-mixing transition amplitude $|\lambda_{a_0}|=
0.60~{\rm GeV^2}$. This is understood from the suppression effect of 
the OZI violation; the inter-mixing is OZI 1st order suppression and the 
intra-mixing is OZI 2nd order suppression.
\section{Inter-, intra- and glueball mixing}
We assume that the inter-mixing between the light scalar mesons $N$ and $L=1\ 
q\bar{q}$ mesons $N'$ are represented as
\begin{eqnarray}
&&L^{eff}_{01}=-\lambda_{01}\epsilon^{abc}\epsilon_{dec}N^d_aN'^e_b=
\lambda_{01}({\rm Tr}(NN')-{\rm Tr}(N){\rm Tr}(N'))\nonumber\\
&&\hspace{1cm}=\lambda_{01}[a^+_0a'^-_0+a^-_0a'^+_0+a^0_0a'^0_0+
\kappa^+K^{*-}_0+\kappa^-K^{*+}_0\nonumber\\
&&\hspace{3cm}+\kappa^0\overline{K}^{*0}_0+\overline{\kappa}^0K^{*0}_0-
f_Nf'_N-\sqrt{2}f_Sf'_N-\sqrt{2}f_Nf'_S],
\end{eqnarray}
where it should be noted that the quark configurations of $I=0$ mesons are 
$f'_N=\frac1{\sqrt{2}}(\bar{u}u+\bar{d}d)$ and $f'_S=\bar{s}s$ for $N'$ but 
$f_N=\frac1{\sqrt{2}}(\bar{s}\bar{d}ds+\bar{s}\bar{u}us)$ and $f_S=\bar{u}
\bar{d}ud$ for $N$. These inter-mixing transitions are represented by the 
OZI rule violating term in only 1st order as shown in Fig.~(4). 
\begin{figure}
\vspace*{1.5cm}
\begin{picture}(5,5)
\hspace*{4cm}
\unitlength 0.1in
\begin{picture}( 24.5000,  6.1000)(  2.4000, -8.0000)
%
\special{pn 13}%
\special{pa 600 200}%
\special{pa 2600 200}%
\special{fp}%
\special{pa 2600 800}%
\special{pa 600 800}%
\special{fp}%
\special{pa 600 600}%
\special{pa 1500 600}%
\special{fp}%
\special{pa 1500 400}%
\special{pa 600 400}%
\special{fp}%
%
\special{pn 13}%
\special{ar 1500 500 100 100  4.7123890 6.2831853}%
\special{ar 1500 500 100 100  0.0000000 1.5707963}%
%
\special{pn 8}%
\special{pa 1300 200}%
\special{pa 1300 800}%
\special{fp}%
\special{pa 1900 800}%
\special{pa 1900 200}%
\special{fp}%
%
\special{pn 8}%
\special{pa 1900 200}%
\special{pa 2610 200}%
\special{pa 2610 800}%
\special{pa 1900 800}%
\special{pa 1900 200}%
\special{ip}%
%
\special{pn 4}%
\special{pa 2600 200}%
\special{pa 2010 790}%
\special{fp}%
\special{pa 2480 200}%
\special{pa 1900 780}%
\special{fp}%
\special{pa 2360 200}%
\special{pa 1900 660}%
\special{fp}%
\special{pa 2240 200}%
\special{pa 1900 540}%
\special{fp}%
\special{pa 2120 200}%
\special{pa 1900 420}%
\special{fp}%
\special{pa 2000 200}%
\special{pa 1900 300}%
\special{fp}%
\special{pa 2610 310}%
\special{pa 2130 790}%
\special{fp}%
\special{pa 2610 430}%
\special{pa 2250 790}%
\special{fp}%
\special{pa 2610 550}%
\special{pa 2370 790}%
\special{fp}%
\special{pa 2610 670}%
\special{pa 2490 790}%
\special{fp}%
%
\special{pn 8}%
\special{pa 590 190}%
\special{pa 1300 190}%
\special{pa 1300 790}%
\special{pa 590 790}%
\special{pa 590 190}%
\special{ip}%
%
\special{pn 4}%
\special{pa 1120 600}%
\special{pa 930 790}%
\special{fp}%
\special{pa 1000 600}%
\special{pa 810 790}%
\special{fp}%
\special{pa 880 600}%
\special{pa 690 790}%
\special{fp}%
\special{pa 760 600}%
\special{pa 590 770}%
\special{fp}%
\special{pa 640 600}%
\special{pa 590 650}%
\special{fp}%
\special{pa 1240 600}%
\special{pa 1050 790}%
\special{fp}%
\special{pa 1300 660}%
\special{pa 1170 790}%
\special{fp}%
%
\special{pn 4}%
\special{pa 1080 400}%
\special{pa 890 590}%
\special{fp}%
\special{pa 960 400}%
\special{pa 770 590}%
\special{fp}%
\special{pa 840 400}%
\special{pa 650 590}%
\special{fp}%
\special{pa 720 400}%
\special{pa 590 530}%
\special{fp}%
\special{pa 1200 400}%
\special{pa 1010 590}%
\special{fp}%
\special{pa 1300 420}%
\special{pa 1130 590}%
\special{fp}%
\special{pa 1300 540}%
\special{pa 1250 590}%
\special{fp}%
%
\special{pn 4}%
\special{pa 1040 200}%
\special{pa 850 390}%
\special{fp}%
\special{pa 920 200}%
\special{pa 730 390}%
\special{fp}%
\special{pa 800 200}%
\special{pa 610 390}%
\special{fp}%
\special{pa 680 200}%
\special{pa 590 290}%
\special{fp}%
\special{pa 1160 200}%
\special{pa 970 390}%
\special{fp}%
\special{pa 1280 200}%
\special{pa 1090 390}%
\special{fp}%
\special{pa 1300 300}%
\special{pa 1210 390}%
\special{fp}%
\special{pn 8}%
\special{pn 13}%
\special{pa 1300 514}%
\special{pa 1306 516}%
\special{pa 1310 520}%
\special{pa 1316 520}%
\special{pa 1320 520}%
\special{pa 1326 520}%
\special{pa 1330 516}%
\special{pa 1336 514}%
\special{pa 1340 508}%
\special{pa 1346 504}%
\special{pa 1350 500}%
\special{pa 1356 494}%
\special{pa 1360 490}%
\special{pa 1366 486}%
\special{pa 1370 482}%
\special{pa 1376 482}%
\special{pa 1380 480}%
\special{pa 1386 482}%
\special{pa 1390 482}%
\special{pa 1396 486}%
\special{pa 1400 490}%
\special{pa 1406 494}%
\special{pa 1410 498}%
\special{pa 1416 504}%
\special{pa 1420 508}%
\special{pa 1426 512}%
\special{pa 1430 516}%
\special{pa 1436 518}%
\special{pa 1440 520}%
\special{pa 1446 520}%
\special{pa 1450 520}%
\special{pa 1456 516}%
\special{pa 1460 514}%
\special{pa 1466 510}%
\special{pa 1470 504}%
\special{pa 1476 500}%
\special{pa 1480 494}%
\special{pa 1486 490}%
\special{pa 1490 486}%
\special{pa 1496 484}%
\special{pa 1500 482}%
\special{pa 1506 480}%
\special{pa 1510 480}%
\special{pa 1516 482}%
\special{pa 1520 486}%
\special{pa 1526 490}%
\special{pa 1530 494}%
\special{pa 1536 498}%
\special{pa 1540 504}%
\special{pa 1546 508}%
\special{pa 1550 512}%
\special{pa 1556 516}%
\special{pa 1560 518}%
\special{pa 1566 520}%
\special{pa 1570 520}%
\special{pa 1576 520}%
\special{pa 1580 518}%
\special{pa 1586 514}%
\special{pa 1590 510}%
\special{pa 1596 506}%
\special{ip}%
\special{pa 1600 500}%
\special{pa 1606 496}%
\special{pa 1610 490}%
\special{pa 1616 486}%
\special{pa 1620 484}%
\special{pa 1626 482}%
\special{pa 1630 480}%
\special{pa 1636 480}%
\special{pa 1640 482}%
\special{pa 1646 484}%
\special{pa 1650 488}%
\special{pa 1656 492}%
\special{pa 1660 498}%
\special{pa 1666 502}%
\special{pa 1670 508}%
\special{pa 1676 512}%
\special{pa 1680 516}%
\special{pa 1686 518}%
\special{pa 1690 520}%
\special{pa 1696 520}%
\special{pa 1700 520}%
\special{pa 1706 518}%
\special{pa 1710 514}%
\special{pa 1716 510}%
\special{pa 1720 506}%
\special{pa 1726 502}%
\special{pa 1730 496}%
\special{pa 1736 492}%
\special{pa 1740 488}%
\special{pa 1746 484}%
\special{pa 1750 482}%
\special{pa 1756 480}%
\special{pa 1760 480}%
\special{pa 1766 482}%
\special{pa 1770 484}%
\special{pa 1776 488}%
\special{pa 1780 492}%
\special{pa 1786 498}%
\special{pa 1790 502}%
\special{pa 1796 506}%
\special{pa 1800 512}%
\special{pa 1806 516}%
\special{pa 1810 518}%
\special{pa 1816 520}%
\special{pa 1820 520}%
\special{pa 1826 520}%
\special{pa 1830 518}%
\special{pa 1836 516}%
\special{pa 1840 512}%
\special{pa 1846 506}%
\special{pa 1850 502}%
\special{pa 1856 496}%
\special{pa 1860 492}%
\special{pa 1866 488}%
\special{pa 1870 484}%
\special{pa 1876 482}%
\special{pa 1880 480}%
\special{pa 1886 480}%
\special{pa 1890 482}%
\special{pa 1896 484}%
\special{pa 1900 488}%
\special{sp}%
%
\special{pn 13}%
\special{pa 1400 200}%
\special{pa 1600 200}%
\special{fp}%
\special{sh 1}%
\special{pa 1600 200}%
\special{pa 1534 180}%
\special{pa 1548 200}%
\special{pa 1534 220}%
\special{pa 1600 200}%
\special{fp}%
\special{pa 1800 800}%
\special{pa 1600 800}%
\special{fp}%
\special{sh 1}%
\special{pa 1600 800}%
\special{pa 1668 820}%
\special{pa 1654 800}%
\special{pa 1668 780}%
\special{pa 1600 800}%
\special{fp}%
\special{pa 1350 600}%
\special{pa 1450 600}%
\special{fp}%
\special{sh 1}%
\special{pa 1450 600}%
\special{pa 1384 580}%
\special{pa 1398 600}%
\special{pa 1384 620}%
\special{pa 1450 600}%
\special{fp}%
\special{pa 1460 400}%
\special{pa 1370 400}%
\special{fp}%
\special{sh 1}%
\special{pa 1370 400}%
\special{pa 1438 420}%
\special{pa 1424 400}%
\special{pa 1438 380}%
\special{pa 1370 400}%
\special{fp}%
\put(2.4000,-5.4000){\makebox(0,0)[lb]{$qq\bar{q}\bar{q}$}}%
\put(26.9000,-5.4000){\makebox(0,0)[lb]{$q\bar{q}$}}%
\end{picture}%
\end{picture}
\caption{Graph for the inter-mixing transition between $qq\bar{q}\bar{q}$ and 
$q\bar{q}$. This graph violats the OZI rule in 1st order.}
\end{figure}
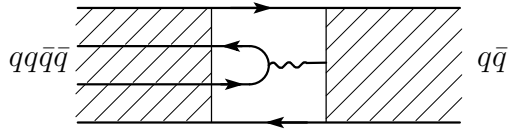
The reason why we assume the transition ${\rm Tr}(NN')
-{\rm Tr}(N){\rm Tr}(N')$ instead of  ${\rm Tr}(NN')$ is that only the 
amplitude ${\rm Tr}(NN')$ contains the transtion $<f_S|f'_S>={<\bar{u}
\bar{d}ud}|\bar{s}s>$ which is the OZI rule violating term in 3rd order and 
then is 
negligible. The strentgthes of transition $\lambda_{01}$ are estimated in 
section 2 as $\lambda_{a_0}=0.60~{\rm GeV^2}$, $\lambda_{K_0}=
0.51~{\rm GeV^2}$ and the reason why the strength is so large and why 
$\lambda_{a_0}$ and $\lambda_{K_0}$ are nearly equal is undrestood easily 
from the Eq. (18).
\par 
In previous section, we considered the intra-mixing among the light scalar 
mesons using the mass matrix represented as 
\setcounter{equation}{15}  
\begin{equation}
\left(\begin{array}{cc}
    m_a^2+2\lambda_0&\sqrt{2}\lambda_0\\
    \sqrt{2}\lambda_0&2m^2_{\kappa}-m^2_a+\lambda_0
    \end{array}\right).
\end{equation}
\setcounter{equation}{18}
Because there is a glueball candidate $f_0(1500)$ \cite{GLUEBALL1} near the 
mass range of the $L=1\ q\bar{q}$ scalar mesons, it is necessary to consider 
the mixing between the glueball and the $L=1\ q\bar{q}$ scalar meson. We 
analyzed this problem in previous paper 
\cite{TESHIMA}. The masses of the $L=1\ q\bar{q}$ mesons before mixing and 
the transition amplitudes among the $L=1\ q\bar{q}$ mesons are described 
by the same effective Lagrangian as the light scalar mesons described 
in Eq.~(12). The strength for the 
transition between the $L=1\ q\bar{q}$ meson and the glueball is described by 
the $\lambda_G$ which corresponds to the graph shown in Fig.~5(a). In our 
previous analysis \cite{TESHIMA}, we used the parameters $\lambda_{GN}$ and 
$\lambda_{GS}$ instead of the parameter $\lambda_{G}$ considering the 
$SU(3)$ violation effect. The glueball mass before mixing is represented by 
the parameter $\lambda_{GG}$ corresponding to the strength of the transition 
between pure glueball and pure glueball as shown in Fig.~5(b). In 
Ref.~\cite{TESHIMA}, we showed this as $m^2_G+\lambda_{GG}$, where 
$\lambda_{GG}$ in the expression represents the contribution from the diagram 
containing a quark loop in two gluon lines in Fig.~5(b). The mass matrix 
representing the intra-mixing among $L=1\ q\bar{q}$ mesons and glueball is 
\begin{equation}
\left(\begin{array}{ccc}
    m_{a'}^2+2\lambda_1&\sqrt{2}\lambda_1&\sqrt{2}\lambda_G\\
    \sqrt{2}\lambda_1&2m^2_{K'}-m^2_{a'}+\lambda_1&\lambda_G\\
    \sqrt{2}\lambda_G&\lambda_G&\lambda_{GG}
    \end{array}\right).
\end{equation} 
\begin{figure}[htb]
\vspace*{2cm}
\begin{picture}(10,4)
\hspace*{1.5cm}\input{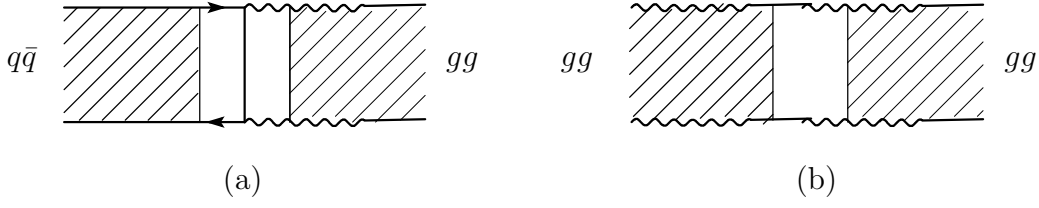}
\end{picture}
\caption{Graphs containing the glueball contributions. Fig.~(a) represents 
the transition between $q\bar{q}$ meson and glueball $gg$. Fig.~(b) represents 
the pure glueball-pure glueball transition.}
\end{figure}
\par 
We consider the overall mixing containing the inter-, intra- and glueball 
mixing represented by the mixing matrix $O$ and mass matrix $M$ as 
\begin{eqnarray}
&&\left(\begin{array}{c}
   f_0(980)\\
   \sigma(600)\\
   f_0(1370)\\
   f_0(1710)\\
   f_0(1500)
   \end{array}\right)=O
   \left(\begin{array}{c}
   f_N\\
   f_S\\
   f'_{N}\\
   f'_{S}\\
   f_G
   \end{array}\right),\ \ \ \ \ \ OM\,{^tO}=M_D,\\
&&M=\left(\begin{array}{ccccc}
    m_N^2+2\lambda_0&\sqrt{2}\lambda_0&\lambda_{01}&\sqrt{2}\lambda_{01}&0\\
    \sqrt{2}\lambda_0&m^2_{S}+\lambda_0&\sqrt{2}\lambda_{01}&
             0&0\\
    \lambda_{01}&\sqrt{2}\lambda_{01}&m_{N'}^2+2\lambda_1&\sqrt{2}\lambda_1&
             \sqrt{2}\lambda_G\\
    \sqrt{2}\lambda_{01}&0&\sqrt{2}\lambda_1&m^2_{S'}+\lambda_1&
             \lambda_G\\
    0&0&\sqrt{2}\lambda_G&\lambda_G&\lambda_{GG}
    \end{array}\right),\nonumber\\
&&M_D={\rm diag}[m^2_{f_0(980)},\ m^2_{\sigma(600)},\ m^2_{f_0(1370)},
    \ m^2_{f_0(1710)},\ m^2_{f_0(1500)}].\nonumber
\end{eqnarray} 
We estimate the best fit values for $\lambda_{01}$, $\lambda_{0}$, 
$\lambda_{1}$, $\lambda_{G}$ and $\lambda_{GG}$ adopting the following mass 
values of $m_{N}$, $m_{S}$, $m_{N'}$ and $m_{S'}$ estimated in section 2; 
\begin{eqnarray}
&&m_N=m_{\overline{a_0(980)}}=m_{\overline{f_0(980)}}=1271{\rm MeV},\ \
m_{\overline{\kappa(900)}}=1047{\rm MeV},\nonumber\\ 
&&\hspace{1cm}m_S=m_{\overline{\sigma(600)}}=760{\rm MeV},\nonumber\\
&&m_{N'}=m_{\overline{a_0(1450)}}=m_{\overline{f_0(1370)}}=1236{\rm MeV},\ \ 
m_{\overline{K^*_0(1430)}}=1307{\rm MeV},\nonumber\\
&&\hspace{1cm}m_{S'}=m_{\overline{f_0(1710)}}=374{\rm MeV},
\end{eqnarray} 
taking the least $\chi^2$ defined by  $\sum_a({m_{a}-m_{a_0}})^2/
{\Delta m_a}^2$, where $m_a$ and ${\Delta m_a}$ represent the experimental 
mass values and mass errors of the scalar meson $a$ and $m_{a_0}$ represents 
the value of the mass estimated. Estimated values of
 $\lambda_{01}$, $\lambda_{0}$, $\lambda_{1}$, 
$\lambda_{G}$ and $\lambda_{GG}$, and $m_{f_0(980)}$, $m_{\sigma(600)}$, 
$m_{f_0(1370)}$, $m_{f_0(1710)}$ and $m_{f_0(1500)}$ are as follows:
\begin{eqnarray}
&&\lambda_{01}=0.51{\rm GeV^2},\ \ \lambda_{0}=0.05{\rm GeV^2},\ \ 
\lambda_{1}=0.05{\rm GeV^2},\nonumber\\ 
&&\lambda_{G}=0.26{\rm GeV^2},\ \ \lambda_{GG}=1.53{\rm GeV^2},\nonumber\\
&&m_{f_0(980)}=0.981(0.980/0.01){\rm GeV},\ \ m_{\sigma(600)}=
0.455(0.600/0.10){\rm GeV},\nonumber\\
&&m_{f_0(1370)}=1.376(1.350/0.05){\rm GeV},\ \ m_{f_0(1710)}=
1.715(1.715/0.007){\rm GeV},\nonumber\\ 
&&m_{f_0(1500)}=1.499(1.500/0.01){\rm GeV},
\end{eqnarray} 
where values in parentheses are (the experimental mass/the experimental 
error) of scalar mesons. Mixing matrix is 
obtained for estimated values of $\lambda$'s, 
\begin{equation}
\left(\begin{array}{c}
    f_0(980)\\
    \sigma(600)\\
    f_0(1370)\\
    f_0(1710)\\
    f_0(1500)
    \end{array}\right)=
    \left(\begin{array}{ccccc}
     0.7129&-0.3282&-0.2223&-0.5548&0.1640\\
     0.1605&0.8402&-0.5056&-0.0604&0.0945\\
     0.0625&0.4027&0.7000&-0.5191&-0.2729\\
     0.5024&0.1550&0.4481& 0.5221& 0.5002\\
     -0.4580&0.0085&0.0639&-0.3828&0.7997
     \end{array}\right)
    \left(\begin{array}{c}
    f_N\\
    f_S\\
    f_{N'}\\
    f_{S'}\\
    f_G
    \end{array}\right).
\end{equation}
\par
The fact that the estimated $\lambda$'s have a character $\lambda_0\sim
\lambda_1\ll\lambda_G\sim\lambda_{01}$ is consistent with the expected 
feature which comes from the OZI-suppressed character. We estimated the 
$\chi^2=2.40$ for this case. On the other hand, when we adopt the inter-mixing 
expressed by the effective Lagrangian ${\rm Tr}(NN')$ instead of ${\rm Tr}
(NN')-{\rm Tr}(N){\rm Tr}(N')$, we get the $\chi^2$ values as $\chi^2=249.9$, 
which cannot be accepted. If we do not include the glueball into the overall 
mixing, the $\chi^2$ value is $\chi^2=3.92$, then the glueball mixing with 
other scalar mesons is preferred to no glueball mixing in the scalar meson 
spectrum. From the estimated mixing parameters (23), we are able to predict 
the decay ratios and decay widths for the scalar meson decays to two 
pseudoscalar mesons $\phi$'s using the coupling among them as
\begin{eqnarray}
&&\varepsilon^{abc}\varepsilon_{def}N^d_a\phi^e_b\phi^f_c\ \ {\rm for}\ 
    N\phi\phi\ {\rm coupling,}\ \cite{SCHECHTER}\nonumber\\
&&N'^b_a(\phi^c_b\phi^a_c+\phi^a_c\phi^c_b)\ \ {\rm for}\ N'\phi\phi\ 
    {\rm coupling.}
\end{eqnarray}   
Using these couplings, we will analyze the decay problem in the next work. 
Here, we want to restrict ourselves to comment that the mixing parameters 
estimated predict that the $f_0(980)$ has a large $s\bar{s}$ like character 
and the $\sigma(600)$ does very little $s\bar{s}$ like character as expected 
from the experiment. 
\section{Conclusion}
Following the re-establishment of the $\sigma(600)$ and the $\kappa(900)$, the 
light scalar mesons $a_0(980)$ and $f_0(980)$ together with $\sigma(600)$ and 
$\kappa(900)$ are considered as the scalar nonet and on the other hand, 
the high mass scalar mesons $a_0(1450)$, $K^*_0(1430)$, $f_0(1370)$ and 
$f_0(1710)$ are turned out to be considered as the $L=1\ q\bar{q}$ scalar 
nonet. However, the masses of the $L=1\ q\bar{q}$ high mass scalar nonet are 
very large compared to other $L=1\ q\bar{q}$ $1^{++}$ and $2^{++}$ mesons. 
We assumed that the high mass of the $L=1\ q\bar{q}$ scalar nonet is caused 
by the mixing with the light scalar nonet. 
\par
In section 2, we estimated the masses of the $L=1\ q\bar{q}$ scalar nonet $N'$ 
and light scalar nonet $N$ before mixing  as 
$m_{\overline{a_0(1450)}}=m_{\overline{f_0(1370)}}=1236{\rm MeV}$, 
$m_{\overline{K^*_0(1430)}}=1307{\rm MeV}$, 
$m_{\overline{f_0(1710)}}=1374{\rm MeV}$,  
$m_{\overline{a_0(980)}}=m_{\overline{f_0(980)}}=1271{\rm MeV}$, 
$m_{\overline{\kappa(900)}}=1047{\rm MeV}$, 
$m_{\overline{\sigma(600)}}=760{\rm MeV}$, using the well-known mass relation 
$m^2(2^{++})-m^2(1^{++})=2(m^2(1^{++})-m^2(0^{++}))$. The strength of 
inter-mixing are estimated from the $a_0(1450)$-$a_0(980)$ 
and $K^*_0(1430)$-$\kappa(900)$ mixings as $\lambda_{a_0}=0.60{\rm GeV^2}$ and 
$\lambda_{K_0}=0.51{\rm GeV^2}$. These are very large and nearly equal. 
These characters are recognized from the OZI-rule and 
the effective Lagrangian of the transition between $N'$ and $N$ discussed 
in section 4. 
\par
In section 3, the structure of the light scalar mesons $N$ was discussed. 
From the consideration of the mass spectrum of $m_{a_0(980)}\sim m_{f_0(980)}>
m_{\kappa(900)}>m_{\sigma(600)}$ and the $s\bar{s}$-like character of 
$f_0(980)$, we concluded that the configuration of the light scalar $I=0$ 
mesons are $f_0(980)\sim \frac1{\sqrt{2}}(\bar{s}\bar{d}ds+\bar{s}\bar{u}us)$ 
and $\sigma(600)\sim \bar{u}\bar{d}ud$. We considered the intra-mixing using 
the effective Lagrangian expressed as $L^{eff}=-a{\rm Tr}(NN)-b{\rm Tr}
(NNM)-\frac12\lambda{\rm Tr}(N){\rm Tr}(N)$, in which the last term 
corresponds to the 2nd order OZI-suppression term. The mixing among the $L=1\ 
q\bar{q}$ nonet is expressed by the same effective Lagrangian and the glueball 
mixing is also considered there. We consider that the $f_0(1500)$ is the most 
probable candidate of the glueball. 
\par
In section 4, we considered the overall inter- and intra-mixing among the 
light 
scalar nonet $N$, $L=1\ q\bar{q}$ nonet $N'$ and glueball $f_G$. The effective 
Lagrangian of the inter-mixing is assumed as $L^{eff}_{01}=\lambda_{01}
({\rm Tr}(NN')-{\rm Tr}(N){\rm Tr}(N'))$. From the best fit analysis of 
$\chi^2$, we obtained the results: $\lambda_{01}=0.51~{\rm GeV^2}$, 
$\lambda_{0}=0.05~{\rm GeV^2},$ $\lambda_{1}=0.05~{\rm GeV^2}$, $\lambda_{G}=
0.26~{\rm GeV^2}$, $\lambda_{GG}=1.53~{\rm GeV^2}$, $m_{f_0(980)}=0.981~
{\rm GeV}$, $m_{\sigma(600)}=0.455~{\rm GeV}$, $m_{f_0(1370)}=1.376~
{\rm GeV}$, $m_{f_0(1710)}=1.715~{\rm GeV}$ and $m_{f_0(1500)}=1.499~
{\rm GeV}$. Obtained mixing parameters are also consistent with the 
experimental characters.

\end{document}